\documentclass{texstyle}

\usepackage{graphicx}
\usepackage{amsthm}
\usepackage[colorlinks=true, allcolors=blue]{hyperref}
\usepackage{booktabs}

\usepackage{amsmath,amsfonts,amssymb}
\usepackage{bbm}
\usepackage{bm}

\usepackage{natbib}

\usepackage[
ragged
]{sidecap}   
\sidecaptionvpos{figure}{t} 

\newcommand{\R}{ {\mathbb{R}} }
\newcommand{\E}{ {\mathbb{E}} }

\newcommand{\new}{ {\textsc{new}} }

\newcommand{\diag}{ {\text{diag}} }

\newcommand{\br}{ {\bf r} }
\newcommand{\bt}{ {\bf t} }

\newcommand{\bx}{ {\bf x} }

\newcommand{\bS}{ {\bf S} }

\newcommand{\by}{ {\bf y} }
\newcommand{\bX}{ {\bf X} }

\newcommand{\bv}{ {\bf v} }

\newcommand{\bI}{ {\bf I} }
\newcommand{\bQ}{ {\bf Q} }
\newcommand{\bV}{ {\bf V} }

\newcommand{\bw}{ {\bf w} }

\newcommand{\bK}{ {\bf K} }

\newcommand{\bbeta}{ {\boldsymbol \beta} }

\newcommand{\bOmega}{ {\boldsymbol \Omega} }
\newcommand{\bomega}{ {\boldsymbol \omega} }
\newcommand{\balpha}{ {\boldsymbol \alpha} }
\newcommand{\bxi}{ {\boldsymbol \xi} }

\title{Efficient computation of predictive probabilities in probit models via expectation propagation}

\author{Augusto Fasano\ist{1}
, Niccolò Anceschi\ist{2}
, Beatrice Franzolini\ist{3}%
\ \\and Giovanni Rebaudo\ist{1}$^,$\ist{4}
\institute{1}{Collegio Carlo Alberto, Turin, IT ({\tt augusto.fasano@carloalberto.org})}
\institute{2}{Duke University, Durham, USA ({\tt niccolo.anceschi@duke.edu})}
\institute{3}{
A*STAR, 
Singapore, SG ({\tt beatricef@sics.a-star.edu.sg})}
\institute{4}{University of Turin, Turin, IT ({\tt giovanni.rebaudo@unito.it})}
}

\begin{document}

\maketitle



\vspace*{-0.5cm}
\abstract{Abstract}{Binary regression models represent a popular model-based approach for binary classification.
In the Bayesian framework, computational challenges in the form of the posterior distribution motivate still-ongoing fruitful research.
Here, we focus on the computation of predictive probabilities in Bayesian probit models via expectation propagation (\textsc{ep}).
Leveraging more general results in recent literature, we show that such predictive probabilities admit a closed-form expression. 
Improvements over state-of-the-art approaches are shown in a simulation study.
}
\smallskip
{probit model, expectation propagation, Bayesian inference, extended multivariate skew-normal distribution}

%
%
\section{Introduction}
\label{sec:1}
Binary regression models represent a default model-based approach for binary classification.
Although the theory in the frequentist setting is well established, flourishing research is still ongoing in the Bayesian framework, where such models are also used as benchmarks for posterior computations \citep{chopin2017leave}.
\begin{samepage}
\end{samepage}
Here, we focus on the approximation of  predictive probabilities via expectation propagation (\textsc{ep}) in the Bayesian probit model     \vspace*{-0.2cm}
\begin{equation}
\label{eq:1}
        y_i\mid \bbeta 
        \overset{ind}{\sim} \textsc{Bern}\left(\Phi\left(\bx_i^\intercal \bbeta\right) \right), \, i=1,\ldots,n; 
        \quad 
    \bbeta 
    \sim \textsc{N}_p(\boldsymbol{0},\nu^2 \bI_p),
    \vspace*{-0.2cm}
\end{equation}
with $\bbeta\in \R^p$ the unknown vector of parameters, $\bx_i\in \R^p$ the covariate vector associated with observation $i$ and $\bI_p$ the identity matrix of dimension~$p$.
$\Phi(t)$ denotes the cumulative distribution function of a standard Gaussian random variable evaluated at $t$ and
$\phi_p(\bt,\bS)$ will denote the density of a $p$-variate Gaussian random variable with mean $\boldsymbol{0}$ and covariance matrix $\bS$, evaluated at $\bt$.

We show that the \textsc{ep} approximate predictive probabilities admit a closed-form expression in terms of the output parameters returned by the \textsc{ep} routine.
Such parameters can be obtained at per-iteration cost of $\mathcal{O}(pn\cdot\min\{p,n\})$, as shown in \citet{anceschi2023bayesian} for a broad class of models and derived in full detail for the probit model in \citet{fasano2023efficient}.

\section{Expectation Propagation (EP) review}
\label{sec:2}
Adapting  more general results derived in \citet{anceschi2023bayesian}, \citet{fasano2023efficient} showed that, calling $\by=(y_1,\ldots,y_n)$, the \textsc{ep} approximation $q(\bbeta) \propto \prod_{i=0}^nq_i(\bbeta)$ of the posterior distribution $p(\bbeta\mid \by)$ for model \eqref{eq:1} can be  obtained by leveraging on extended skew-normal (\textsc{sn}) distributions \citep{azzalini2013skew}.
Except for $q_0(\bbeta)$, which is fixed equal to the prior $p(\bbeta)$, we take $q_i(\bbeta)=\phi_p\left(\bbeta-\bQ_i^{-1}\br_i,\bQ_i^{-1}\right)$, $i=1,\ldots,n$, with the optimal $\br_i$'s and $\bQ_i$'s to be obtained via the \textsc{ep} routine.
Consequently, calling $\br_0 = \boldsymbol{0}$ and $\bQ_0=\nu^{-2}\bI_p$, one gets $q(\bbeta) = \phi_p(\bbeta-\bQ^{-1}\br,\bQ^{-1})$, with $\br=\sum_{i=0}^n \br_i$, $\bQ = \sum_{i=0}^n\bQ_i$.
At each \textsc{ep} cycle, the parameters $\br_i$ and $\bQ_i$ of each site $i=1,\ldots,n$ are updated by imposing that the first two moments of the global approximation $q(\bbeta)$ match the ones of the hybrid distribution \vspace*{-0.1cm}
\begin{equation}
\label{eq:2}
	h_i(\bbeta) \propto p(y_i\mid \bbeta) \prod_{j\ne i} q_j(\bbeta) =
	\Phi((2y_i-1)\bx_i^\intercal\bbeta)\prod_{j\ne i} q_j(\bbeta).
 \vspace*{-0.2cm}
\end{equation}
This is immediate after noticing that \eqref{eq:2} coincides with the kernel of a multivariate extended skew-normal distribution $\textsc{sn}_p(\bxi_i,\bOmega_i,\balpha_i,\tau_i)$, with
\vspace*{-0.1cm}
\begin{equation*}
    \bxi_i = \bQ_{-i}^{-1}\br_{-i},\ \ 
	\bOmega_i = \bQ_{-i}^{-1},\ \ 
    \balpha_i = (2y_i-1)\bomega_i\bx_i,\ \ 
	\tau_i = (2y_i-1)(1+\bx_i^\intercal\bOmega_i\bx_i)^{-1/2}\bx_i^\intercal\bxi_i,
 \vspace*{-0.1cm}
\end{equation*}
where $\bQ_{-i}=\sum_{j\ne i} \bQ_j$, $\br_{-i}=\sum_{j\ne i}\br_j$ and $\bomega_i=\left[\text{diag}\left(\bOmega_i\right)\right]^{1/2}$.
Combining this with Woodbury's identity, \citet{fasano2023efficient} show that, for $i=1\ldots,n$, the updated quantities $\bQ_i^\new$ and $\br_i^\new$ 
equal $k_i \bx_i \bx_i^\intercal$ and $m_i \bx_i$, respectively,
with $k_i = -\zeta_2(\tau_i)/\left(1 + \bx_i^\intercal\bOmega_i\bx_i + \zeta_2(\tau_i)\bx_i^\intercal\bOmega_i\bx_i\right)$ and $m_i = \zeta_1(\tau_i) s_i + k_i   (\bOmega_i\bx_i)^\intercal \br_{-i} + k_i \zeta_1(\tau_i) s_i \bx_i^\intercal \bOmega_i\bx_i$, having defined $\zeta_1(x) = \phi(x)/\Phi(x)$, $\zeta_2(x)=-\zeta_1(x)^2-x\zeta_1(x)$ and $s_i=(2y_i-1)(1+\bx_i^\intercal\bOmega_i\bx_i)^{-1/2}$.
These results, combined with the efficient computation of $\bOmega_i$ and update of the covariance matrix $\bQ^{-1}$ of the Gaussian approximation $q(\bbeta)$, lead to an implementation of \textsc{ep} having a cost per iteration $\mathcal{O}(p^2 n)$.
When $p$ is large, and especially when $p>n$, \textsc{ep} can be implemented at $\mathcal{O}(p n^2)$ cost per iteration by storing and updating only the $p$-dimensional vectors $\bw_i = \bOmega_i \bx_i = \bQ_{-i}^{-1}\bx_i$ and $\bv_i = \bQ^{-1}\bx_i$, $i=1,\ldots,n$.
Eventually, one can compute the full \textsc{ep} covariance matrix as \vspace*{-0.1cm}
\begin{equation}
\label{eq:3}
    \bQ^{-1}=\nu^2\bI_p-\nu^2\bV\bK\bX,\vspace*{-0.1cm}
\end{equation}
where $\bV=[\bv_1,\dots,\bv_n]$, $\bX=[\bx_1,\dots,\bx_n]^\intercal$ and $\bK = \diag(k_1,\ldots,k_n)$.

\section{Closed-form EP predictive probabilities}
\label{sec:3}
One of the advantages of the Gaussian approximation provided by \textsc{ep} is that it results in a simple closed-form expression for the approximate predictive probability of observing $y_\new=1$ for a new statistical unit having covariate vector $\bx_\new$, namely $\Pr_{\textsc{ep}}[y_\new=1\mid \by]$.
Indeed, calling $\bxi_{\textsc{ep}}=\bQ^{-1}\br$ and $\bOmega_{\textsc{ep}}=\bQ^{-1}$ so that $q(\bbeta)=\phi_p\left(\bbeta-\bxi_\textsc{ep},\bOmega_\textsc{ep}\right)$, it holds\vspace*{-0.2cm}
\begin{equation}
\label{eq:4}
    \Pr{}_{\textsc{ep}}[y_\new=1\mid \by] 
    =\E_{q(\bbeta)}\big[\Phi\big(\bx_\new^\intercal \bbeta \big)\big]
    =\Phi\big( \big(1 + u\big)^{-1/2} \bx_\new^\intercal \bxi_\textsc{ep}\big),\vspace*{-0.2cm}
\end{equation}
where $u=\bx_\new^\intercal \bOmega_{\textsc{ep}}\bx_\new$ and the last equality in \eqref{eq:4} follows by Lemma 7.1 in \citet{azzalini2013skew}.
The only computationally relevant part in \eqref{eq:4} is the computation of the quadratic form $u$.
However, when $p<n$, $\bOmega_{\textsc{ep}}$ is directly returned by the algorithm, and $u$ can be computed at cost $\mathcal{O}(p^2)$.
On the other hand, when $p>n$ (or in general when $p$ is large), this direct computation can be avoided since, by \eqref{eq:3}, $u=\nu^2\left[\bx_\new^\intercal \bx_\new - \big(\bV^\intercal\bx_\new\right)^\intercal \bK \left(\bX\bx_\new\right)\big]$, computable at cost $\mathcal{O}(pn)$.
Thus, Equation \eqref{eq:4} provides an efficient closed-form approximation of the exact predictive probability \mbox{$\Pr[y_\new=1\mid \by] $}, which can be computed at cost $\mathcal{O}(p\cdot\min\{p,n\})$ from the \textsc{ep} parameters.

\section{Simulation study}
\label{sec:4}
\begin{SCfigure}
\includegraphics[scale=.35]{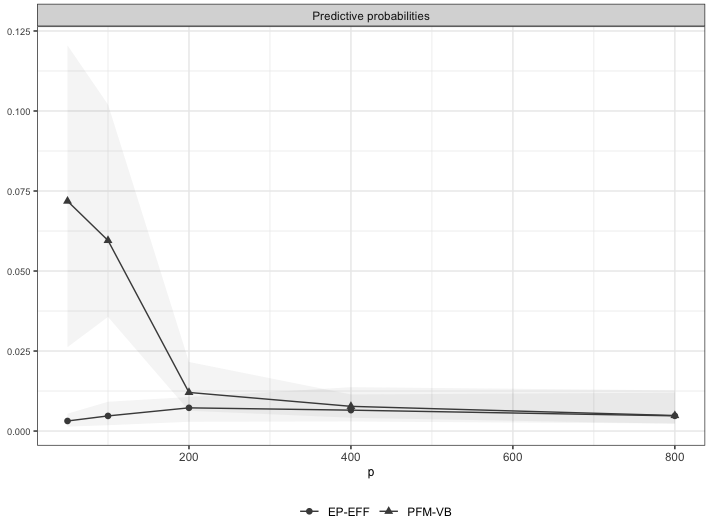}
%
%
\caption{For varying $p$, median absolute difference between the $\tilde{n}=50$  predictive probabilities resulting from $2000$ i.i.d.\ samples and the ones arising from \textsc{ep} and \textsc{pfm-vb} for probit regression with $n=100$ and $\nu^2=25$.
    Grey areas denote the first and third quartiles.}
\label{fig:1}       
\end{SCfigure}

We show with a simulation study the advantages of combining the efficient \textsc{ep} implementation presented in \citet{fasano2023efficient} with the efficient computation of the predictive probabilities presented in Section \ref{sec:3}.
Fixing $n=100$ and $\nu^2=25$, we compute the predictive probabilities for $\tilde{n}=50$ test units in five different scenarios with synthetic data, for $p=50,100,200,400$ and $800$.
We compare the approximate predictive probabilities obtained with \textsc{ep} and with the partially-factorized variational approximation (\textsc{pfm-vb}) (Equation (9) in \citet{fasano2022scalable}) with the ones arising from a Monte Carlo approximation exploiting i.i.d.\ samples from the posterior \citep{durante2019conjugate}.
Figure \ref{fig:1} shows that \textsc{ep} can achieve superior accuracy for $p<2n$, while in the other settings they provide comparable results.
The \textsc{ep} running time ranges from $0.02$ to $0.12$ seconds, while for \textsc{pfm-vb} it ranges from $0.13$ to $0.23$.
The slightly higher cost of \textsc{pfm-vb} is because, after convergence, the computation of predictive probabilities requires a sampling step that takes approximately 0.12 seconds.
To conclude, the results presented in this work make the computation of \textsc{ep} approximate predictive probabilities feasible in settings where currently-available implementations are computationally impractical.
Considering $p=800$ for illustration, the function \texttt{EPprobit} from the \texttt{R} package \texttt{EPGLM}, requires $140$ seconds, about $1000$ times slower than the efficient implementation presented here.
Code is available at \href{https://github.com/augustofasano/EPprobit-SN}{https://github.com/augustofasano/EPprobit-SN}.

%

\vspace*{-0.2cm}
\bibliographystyle{authordate3}
\bibliography{biblio}
\end{document}